\documentclass[12pt, a4paper]{article}
\usepackage{fullpage}
\usepackage{amsmath}
\usepackage{amsfonts}
\usepackage{amssymb}
\usepackage{graphicx, rotating}
\usepackage{epstopdf}
\usepackage{epsfig}
\usepackage{latexsym}
\usepackage{graphicx}
\usepackage{color}
\usepackage{amsmath,amssymb}
\usepackage{cite}
\usepackage{slashed}
\usepackage{hyperref}
\usepackage{comment}
\usepackage{multirow}
\usepackage{hhline}
\usepackage{amsmath}
\usepackage{bm}
\usepackage[utf8]{inputenc} 
\hypersetup{colorlinks, citecolor=bluscuro, linkcolor=black, urlcolor=bluscuro}
\definecolor{rossos}{cmyk}{0,1,1,0.55}
\definecolor{bluscuro}{rgb}{0.15, 0.2, .85}
\definecolor{bluchiaro}{cmyk}{1,.3,0.,0.1}
\newcommand{\eq}[1]{Eq.~(\ref{#1})}
\newcommand{\lag}{\mathcal{L}}

\newcommand{\nn}{\nonumber}

\newcommand{\be}{\begin{equation}}
\newcommand{\ee}{\end{equation}}
\newcommand{\bea}{\begin{eqnarray}}
\newcommand{\eea}{\end{eqnarray}}
\newcommand{\bc}{\begin{center}}
\newcommand{\ec}{\end{center}}
\newcommand{\sfrac}[2]{#1/#2}

\newcommand{\op}{\mathcal{O}}

\def\lra#1{\overset{\text{\scriptsize$\leftrightarrow$}}{#1}}
\begin{document}

\vspace*{-2cm}
\begin{flushright}
\today
\end{flushright}

\begin{center}
\vspace*{15mm}

\vspace{1cm}
{\large \bf 
Improved BSM Sensitivity in Diboson Processes \\at Linear Colliders 
} \\
\vspace{1.4cm}

{Brian Henning, Davide Maria Lombardo and Francesco Riva}

 \vspace*{.5cm} 
 {\it D\'epartment de Physique Th\'eorique, Universit\'e de Gen\`eve,
24 quai Ernest-Ansermet, 1211 Gen\`eve 4, Switzerland }

\vspace*{.2cm} 

\end{center}

\vspace*{10mm} 
\begin{abstract}\noindent\normalsize

  We study $W^+W^-$ and $Zh$ final states at future linear $e^+e^-$ colliders; designing analyses specific to the various final state polarizations allows us to target specific beyond the Standard Model (BSM) effects, parametrized in the form of dimension-6 operators. We find that CLIC can access effects roughly an order of magnitude smaller than HL-LHC or ILC, and two orders of magnitude smaller than LEP. These results are interpreted in the context of well-motivated BSM scenarios---at weak and strong coupling---where we expect correlated effects in Drell-Yann processes. 
 The latter turn out to  have better discovery potential, although the diboson processes provide additional discriminating power, potentially furnishing a way to measure the spin and coupling of BSM states.

\end{abstract}

\vspace*{3mm}

\newpage

\section{Motivation}

Standard Model (SM) precision tests are at the core of present and future collider programs. While interesting as confirmation of our unprecedented control  of SM computations, they serve an exciting additional purpose: a way to search for new structure lurking beyond the SM (BSM). Indeed, heavy dynamics---beyond the direct reach of colliders---leave an imprint on lower energy processes, in the form of  deformations of SM interactions. These  can be described by an Effective Field Theory (EFT), which parametrises the most general deviations from the SM and, at the same time,  captures the effects of general, heavy BSM dynamics.

The leading such effects, associated with dimension-6 operators in the effective Lagrangian, in general behave schematically as $\sigma\sim\sigma_{SM}(1+ E^2/\Lambda^2)$, with $E$ a characteristic energy scale of the process and $\Lambda$ the physical scale associated with the EFT operator. This suggests two modes of exploration: \emph{i)} on a SM resonance---such as the $Z$-pole studied at LEP1---where $\sigma_{SM}$ is maximal, statistical uncertainty the smallest, and the experiment becomes sensitive  to tiny departures from the SM, or \emph{ii)} at high-energy, where the BSM effect is larger and  less precision is needed to access effects of a given size. In this article we focus on experiments of the latter type; in particular, we study non-resonant $2\to2$ processes, which are the simplest processes, with the largest cross sections, with access to the high-energy regime.
Processes with more particles can be interesting to test operators with more legs, such as those that modify Higgs couplings \cite{Henning:2018kys}.

Particularly interesting for the BSM discovery potential they offer are diboson $VV^\prime$ and $Vh$ ($V^{(\prime)}=W,Z$) final states. Indeed, new dynamics in the Higgs sector---such as Higgs compositeness~\cite{Kaplan:1983fs,Kaplan:1983sm,Contino:2003ve,Agashe:2004rs,Giudice:2007fh}---alter the behaviour of $Vh$ processes \cite{Biekoetter:2014jwa,Banerjee:2018bio,Liu:2019vid} and, according to the equivalence theorem, also enter in processes with longitudinal $VV^\prime$ \cite{Franceschini:2017xkh,Durieux:2017rsg,Grojean:2018dqj}. BSM in the gauge sector \cite{Liu:2016idz,Henning:2014wua} instead affects $VV^\prime$ processes with transverse polarisations. Finally, light fermion substructure, as implied in models of fermion compositeness \cite{Kaplan:1991dc,Bellazzini:2017bkb}, can modify the initial light quark or lepton current, although this  option seems to be disfavoured by tests of flavour and CP violation \cite{KerenZur:2012fr}.

Diboson processes happen to be very interesting also for the experimental challenges they pose. Processes involving the transverse polarizations suffer from suppressed SM-BSM interference; this is because the tree-level, high-energy diboson helicity structure in the SM ($\pm,\mp$) differs from ($\pm,\pm$) implied by the leading BSM effects~\cite{Azatov:2016sqh}. However, by utilizing exclusive information in the form of differential distributions of the azimuthal angles of the $V$ boson decay planes, dedicated experiments can bear the interference information and overcome this problem~\cite{Panico:2017frx,Azatov:2017kzw}. 
When new physics is in the longitudinal polarizations, BSM searches as precision tests are also challenging. In the SM, the unpolarized cross section is dominated by the transverse-transverse components while the longitudinals are small. Therefore, even though the SM and BSM do interfere, the dominant SM contribution acts as an irreducible background, thereby reducing the sensitivity of the experiment. 
As we will see, beam polarization can play a crucial role in this context as it can substantially reduce the transverse component. 

Lepton colliders 
offer an ideal environment to study these dedicated experiments~\cite{Ellis:2015sca,Ellis:2017kfi,DiVita:2017vrr}: beside providing an ideal context for precision studies, these machines also present a number of qualitative differences with respect to hadron colliders, such as the possibility of beam polarisation  and, in principle, the knowledge of the collision center-of-mass energy.
In this article we discuss and compare ILC and CLIC capabilities, which offer the best prospects to explore the high-energy regime. We establish the added value of dedicated BSM searches for EFT dimension-6 operators, and discuss the reach of different experiments, at different colliders. In particular, we compare $VV$ against $Vh$ studies when new physics enters in the Higgs/longitudinals sectors, and put these in perspective with precise $Z$-pole measurements at LEP and future high-intensity  circular-collider facilities, which are capable of testing the same physics with high precision. We will see that $Vh$-processes lead to the furthest reach on BSM, both at ILC and CLIC.


In section \ref{sec:BSM} we discuss different weakly and strongly coupled BSM scenarios that can induce energy-growing effects in diboson processes and identify interesting patterns of EFT Wilson coefficients stemming from well-motivated microscopic assumptions. In section \ref{sec:reach}, based on our knowledge of  SM and BSM interplay, we design dedicated collider analyses for the transverse $W^+_TW^-_T$  and longitudinal $W^+_LW^-_L$ final states, as well as associated Higgs production $Zh$. We discuss our results in section \ref{sec:interp}.

Analyses of this type have already been carried out in the context of the LHC, see e.g. Refs.~\cite{Banerjee:2018bio,Azatov:2017kzw,Farina:2016rws}. Preliminary versions of parts of this analysis can be found in the CLIC Potential for New Physics report \cite{deBlas:2018mhx} and in \cite{Brooijmans:2018xbu}. Moreover, Ref.~\cite{deBlas:2019wgy}, appeared recently, has results which slightly overlap with our analysis of longitudinal polarizations.

\section{BSM perspective}\label{sec:BSM}
We are interested in the leading BSM effects, as captured by the most relevant BSM operators in an EFT parametrization 
\begin{equation}
\lag^{eff}=\lag^{SM}+\sum_ic_i\frac{\op_i}{\Lambda^2}+\cdots,
\end{equation}
with $\op_i$ dimension-6 operators \cite{Buchmuller:1985jz,Grzadkowski:2010es,Giudice:2007fh}. We will work in the SILH basis of Ref.~\cite{Giudice:2007fh} which captures efficiently the effects of UV universal theories, where the new dynamics couples principally to the SM bosons.
Different microscopic dynamics in the UV (at $E\gtrsim \Lambda$) translate at low energy into different patterns for the coefficients; in what follows we will describe such patterns in the transverse and longitudinal sectors.

\paragraph{Transverse BSM.}
Heavy new physics that couples to the electroweak gauge interactions, \textit{i.e.} heavy particles charged under $SU(2)_L\times U(1)_Y$, at low-energy induce effects that can be captured by\footnote{Here we focus on CP-even BSM physics, as CP-odd effects of this type are well constrained by measurements of anomalous dipole moments~\cite{Panico:2018hal}.}
\begin{equation}\label{optrans}
{\cal O}_{2B}=\frac 12 \left(\partial^\nu B_{\mu\nu}\right)^2\,,\quad {\cal O}_{2W}=\frac 12 \left(D^\nu W^a_{\mu\nu}\right)^2\,,\quad {\cal O}_{3W}= \frac{1}{3!} g\epsilon_{abc}W^{a\, \nu}_{\mu}W^{b}_{\nu\rho}W^{c\, \rho\mu}\,.
\end{equation}
The first two operators in \eq{optrans} modify the $W^\pm$ and $Z$ boson propagators, where they correspond to the  electroweak  oblique parameters $Y$ and $W$ \cite{Barbieri:2004qk}; their effects can be searched for very efficiently in Drell-Yann (DY) processes \cite{Barbieri:2004qk,Farina:2016rws}.
${\cal O}_{3W}$ enters, for instance, in processes with transverse diboson final states at colliders. In this context, ${\cal O}_{3W}$ is often presented as the anomalous trilinear gauge coupling (TGC), associated with the parameters $\lambda_{\gamma},\lambda_{Z}$ of Ref. \cite{Hagiwara:1986vm}, to which it translates as  $ \lambda_{Z}=\lambda_\gamma=- c_{3W}\sfrac{m_W^2}{\Lambda^2}$ (for concrete reference, \(\lambda_{\gamma} = -0.006c_{3W}(1 \text{ TeV}/\Lambda)^2\)). At leading order, 
\begin{equation}
  \frac{W}{\lambda_{\gamma}} = - \frac{c_{2W}}{c_{3W}}.
\end{equation}


Different BSM scenarios are characterized by different relative importances of ${\cal O}_{2W,2B}$ versus ${\cal O}_{3W}$. 
We focus here on  ${\cal O}_{2W,3W}$ that are typically generated together. 
When the BSM physics is weakly coupled, we have a UV Lagrangian from which we can compute the Wilson coefficients \ directly. This weak coupling assumption on the microphysics implies that we can restrict attention to renormalizable interactions in the UV Lagrangian. 
In this case, \(\mathcal{O}_{3W}\) can only be generated  at one-loop or beyond~\cite{Arzt:1994gp,deBlas:2017xtg}.\footnote{One can also understand this from helicity amplitudes. \(\mathcal{O}_{3W}\) gives a three-point amplitude between gauge bosons of all the same helicity---\textit{i.e.} \(\lambda_{\gamma}\) measures the \(+++\) and \(---\) coupling of EW gauge bosons. Moreover, in renormalizable theories, tree-level amplitudes of all the same helicity vanish. As we restricted ourselves to renormalizable interactions in the UV Lagrangian, this implies that we cannot generate \(\mathcal{O}_{3W}\) at tree-level.}
\(\mathcal{O}_{2W}\) also typically gets generated starting at 1-loop, unless 
heavy particles couple linearly to the \(SU(2)_L\) current \(J_{W,\mu}^a\)~\cite{Arzt:1994gp,Giudice:2007fh}, in which case it can be generated at tree-level.
Lorentz invariance, gauge invariance, and linearity then imply the particle must be a massive vector in the adjoint representation of \(SU(2)_L\). Thus, unless the microphysics contains an EW triplet vector, \(\mathcal{O}_{2W}\) is generated at one-loop or beyond.

The one-loop contribution to \(\mathcal{O}_{2W,3W}\)  depends on the heavy particle's mass, spin, and EW charge; in particular, it is independent of coupling constants in the UV Lagrangian while its dependence on the coupling $g$ is fixed by gauge invariance. Moreover, each particle separately generates these operators, \(C_{2W,3W} = \sum_i C_{2W,3W}^i\) where the sum is over the BSM particles and we define $C_{2W,3W}=c_{2W,3W}/\Lambda^2$. The contributions to \(C_{2W,3W}^i\) from a particle of mass \(M_i\) and in the \(R_i\)th representation of \(SU(2)_L\) are~\cite{Henning:2014wua}
\begin{subequations}\label{eq:C2W3W}
  \begin{align}
    C_{2W}^i &= \frac{1}{(4\pi)^2}\,\frac{1}{M_i^2}\,\frac{g^2}{60}\,\mu(R_i) \, \cdot a_{2W}^i, \\
    C_{3W}^i &= \frac{1}{(4\pi)^2}\,\frac{1}{M_i^2}\,\frac{g^2}{60}\,\mu(R_i) \, \cdot a_{3W}^i,
  \end{align}
\end{subequations}
where \(\mu(R_i)\) is the Dynkin index, \(\text{tr}_{R_i}(T^aT^b) = \mu(R_i)\delta^{ab}\), and \(a_{2W,3W}\) are constants which depend on the spin of the particle, as shown in table~\ref{eq:a2W3W}. Interestingly, the ratio \(C_{2W}^i/C_{3W}^i\) only depends on the spin of the particle, and this value varies wildly from \(+1\) for a scalar to \(-37/3\approx -12\) for a vector. As we discuss in detail in sec.~\ref{sec:interp}, this implies that the ratio \(C_{2W}/C_{3W}\) (equivalently, \(W/\lambda_{\gamma}\)) is a potentially excellent indicator of the spin of the lightest BSM state that generates the operators. 

\begin{table}[ht]
\begin{center}
\begin{tabular}{l|ccc}
    & $a_{2W}$ & $a_{3W}$ & $C^i_{2W}/C^i_{3W}$ \\ \hline
    Real scalar & 1 & 1 &1 \\
     Complex scalar & 2 & 2 &1 \\
    Dirac fermion & 16 & $-4$ &$-4$ \\
    Vector & $-37$ & 3 &$-37/3$\\
\end{tabular}
\end{center}
\caption{\it Coefficients appearing in \eq{eq:C2W3W} and ratio of Wilson coefficients resulting from integrating out  different UV-particles.}
\label{eq:a2W3W}
\end{table}%

One can massage the results of~\cite{Henning:2014wua} to re-express this ratio as
\begin{equation}\label{eq:ratio_reduced}
  \frac{C_{2W}^i}{C_{3W}^i} = 1 - 20 \frac{k^i(j_1,j_2)}{N^i_{\text{dof}}}.
\end{equation}
In this expression, \(N^i_{\text{dof}}\) is the number of \textit{physical} real degrees of freedom of the \(i\)th \textit{particle} (respectively, \(N_{\text{dof}} = 1\), 4, and 3 for a real scalar, Dirac fermion, and a massive vector), while \(k^i(j_1,j_2)\) is the Dynkin index for a \textit{field} in the \((j_1,j_2)\)th representation of \(SO(3,1)\).\footnote{\(\text{tr}_{(j_1,j_2)}\big(\mathcal{J}^{\mu\nu}\mathcal{J}^{\rho\sigma} \big) = k(j_1,j_2)\big(g^{\mu\rho}g^{\nu\sigma} - g^{\mu\sigma}g^{\nu\rho}\big)\). For a gauge field \(k = 2\), for a Dirac fermion \(k= 1\), while \(k=0\) for scalars, Goldstone fields, and ghosts.}
To arrive at this equation, one needs to sum over contributions from all fields used to describe the particle; in particular, for the massive vector, the contributions of Goldstone fields (describing the longitudinal mode) and ghost fields (to cancel unphysical polarizations) are accounted for.
By direct computation we know the above holds for scalars, fermions, and vectors; it would be interesting to understand if it extends to massive higher spin particles.\footnote{The point here is that \(N_{\text{dof}}\) is a physical quantity, but that is not necessarily true of the Dynkin index \(k\). For massive vectors, the Goldstone and ghost fields are scalars and therefore have \(k=0\), which is why eq.~\eqref{eq:ratio_reduced} only needs \(k\) for the gauge field. For massive higher spin particles, the Goldstone and ghost fields can be in non-trivial representations of \(SO(3,1)\). The question one wants to answer is if these contribute in such a way that eq.~\eqref{eq:ratio_reduced} could be expressed in terms of only physical quantities (one speculative guess is that it depends only on \(N_{\text{dof}}\) and the Dynkin index of the \(SO(3)\) \textit{little group} representation to which the particle belongs). We note that even though the question at hand---what is the one-loop contribution from massive higher-spin states---is technically well posed, the regime of validity of the perturbative expansion is narrow: the higher-spin action is itself an effective action with a cutoff, and Ref.~\cite{Bellazzini:2019bzh} showed that a single higher-spin particle cannot be parametrically lighter than this cutoff, so that a weak coupling treatment of isolated higher-spin particles is typically not possible.}


Our discussion has focused on how the ratio \(C_{2W}/C_{3W}\) encodes valuable kinematic information about the spin. But what of the overall size of these coefficients? 
From eq.~\eqref{eq:C2W3W}, we see these coefficients are proportional to \(\mu(R)\), which is the non-abelian analogue of \(Q^2\) (roughly, it's the sum of \(Q^2\) for each state in the representation). For the spin \(j\) representation of \(SU(2)\),
\begin{equation}
  \mu(R_j)=\sum_{i=0}^{2j}(j-i)^2 = \frac{1}{3}j(j+1)(2j+1),
\end{equation}
which scales as \(2j^3/3\) at large \(j\). 
The possibility of a large $\mu(R)$ to compensate for the loop suppression factors in \eq{eq:C2W3W} is important in the discussion of the validity of our EFT parametrization, see e.g. Ref.~\cite{Contino:2016jqw}. Indeed, at $E\sim 2M$ the heavy particles can be produced on shell and the EFT parametrization collapses. From \eq{eq:C2W3W} we see that at that energy, the operators $\op_{2W}$ and $\op_{3W}$ can produce at most a relative effect  $C E^2\sim (10^{-3}\div 10^{-4}) \mu(R) $ with respect to the SM, corresponding to a very precise relative measurement for small $\mu(R)$. Larger \(SU(2)_L\) representations could instead potentially compensate mass scales beyond direct collider production.

This problem, which is a showstopper for hadron collider searches of these effects, has pushed  the development of scenarios where the transverse polarizations can be inherently strongly coupled, wherein \eq{eq:C2W3W} no longer applies. 
In addition to the usual monopole coupling $g$ appearing in the covariant derivative, the transverse polarisations in these scenarios have another coupling $g_*$ characterising dipole- and multipole-type interactions~\cite{Liu:2016idz}.
The latter dominate at high-energy  where they are parametrically enhanced by  $g_*$, resulting in the following estimate for the Wilson coefficients,
\begin{equation}\label{remediosscaling}
c_{2W,2B}\sim O(1)\quad c_{3W}\sim \frac{g_*}{g}.
\end{equation}
Now $c_{3W}$ can be larger than one, and even an experiment with $O(1)$ resolution can test the multipolar interaction hypothesis consistently.

\paragraph{Longitudinals.} 
Concerning new physics in the Longitudinals/Higgs sector, one can identify a large number of operators that potentially contribute. 
For instance, in the SILH basis, the most important energy-growing effects  that enter in diboson processes are captured by
\begin{eqnarray}
{\cal O}_W=\frac{ig}{2}( H^\dagger  \sigma^a \lra {D^\mu} H  )D^\nu  W_{\mu \nu}^a\,,\quad 
{\cal O}_B=\frac{ig'}{2}( H^\dagger  \lra {D^\mu} H  )\partial^\nu  B_{\mu \nu}\,,\label{opslong}\\
{\cal O}_{HW}=i g(D^\mu H)^\dagger\sigma^a(D^\nu H)W^a_{\mu\nu}\,,\quad
{\cal O}_{HB}=i g'(D^\mu H)^\dagger(D^\nu H)B_{\mu\nu}\,,\nn
\end{eqnarray}
These modify the Longitudinals' and Higgs' sectors due to the presence of \(H\) in the operators, as \(H\) contains the physical Higgs \(h\) as well as the Goldstone fields that become the longitudinal polarizations. In perturbative models of spin$\leq 1$ coupled with strength $g_*$,  $c_{W,B}$ can arise at tree-level, while $c_{HW,HB}$ arise at loop-level~\cite{Arzt:1994gp,Giudice:2007fh,Greco:2014aza}:
\begin{equation}
c_{W,B}\sim 1, \quad  c_{HW,HB}\sim \frac{g_*^2}{16\pi^2}.
\end{equation}

As discussed at length in Ref.~\cite{Franceschini:2017xkh,Grojean:2018dqj}, in the high-energy limit of $\bar{\psi} \psi \to V_LV^\prime_L$ processes, fixing the initial state polarization of the fermions determines the unique combination of operators in \eq{opslong} that contribute to the process. This further implies that, in the high-energy regime (\textit{i.e.} the massless limit), these processes do not distinguish between effects that modify the propagator and effects that modify vertices. For the initial state polarizations \(\bar{\psi}_R \psi_L\) and \(\bar{\psi}_L\psi_R\) always involve \(\mathcal{O}_W\) and \(\mathcal{O}_B\). Considering this, together with the fact that these operators can be generated at tree-level, in what follows we will focus our attention on \(\mathcal{O}_W\) and \(\mathcal{O}_B\). 



\section{Collider Reach} \label{sec:reach}
\subsection{Transverse $e^+e^-\to W_T^+W_T^-$}\label{sec:antrans}

BSM effects in amplitudes with transverse $W_T^+W_T^-$ final states are difficult to test due to small interference between SM and BSM amplitudes. As we will see, however, an understanding of angular distributions in the \(W\) decay products provides a way of digging out the BSM signal.

${\cal O}_{3W}$ produces, at tree-level and at high-energy, dominantly $++$ or $--$  helicities in the final states, with amplitudes 
\begin{equation}\label{bsmTT}
({\mathcal{A}}_{\textrm{\tiny{BSM}}})^{+\,+}_L=({\mathcal{A}}_{\textrm{\tiny{BSM}}})^{-\,-}_L\approx c_{3W} g^2\frac{s}{\Lambda^2} \sin\Theta \,,\quad ({\mathcal{A}}_{\textrm{\tiny{BSM}}})^{+\,+}_R\approx ({\mathcal{A}}_{\textrm{\tiny{BSM}}})^{-\,-}_R\approx 0,
\end{equation}
where $\Theta$ is the polar angle, corresponding to the angle between the incoming electron and the outgoing $W^-$, $\sqrt{s}$ the center-of-mass energy, and the subscripts $L$ or $R$ denote the helicity of the incoming electrons, while the supscripts $\pm$ the helicity of the outgoing $W^+$ and $W^-$.  In inclusive $2\to 2$ scattering, the BSM amplitudes in \eq{bsmTT} do not interfere with the SM amplitude ${\cal A}_{SM}$, which is dominated at high-energy by the $+-$ and $-+$ helicities with amplitudes
\begin{align}\label{ampssm}
& ({\cal A}^{-+}_{\textrm{\tiny{SM}}})_{\text{L}}\approx -\dfrac{g^2}{2} \sin\Theta ,\quad\quad
({\cal A}^{+-}_{\textrm{\tiny{SM}}})_{\text{L}}\approx 2g^2  \cos ^4\frac{\Theta}{2}  \csc \Theta ,\quad\quad ({\cal A}^{-+}_{\textrm{\tiny{SM}}})_{\text{R}}\approx ({\cal A}^{+-}_{\textrm{\tiny{SM}}})_{\text{R}}\approx 0.
\end{align}
However, the amplitudes for $e^+e^-\to W^+W^-\to 4\psi$ decaying into fermions do interfere,  proportionally to a function of the azimuthal angles $\varphi^+$ and $\varphi^-$ of the decay planes of the fermion/anti-fermion originating from the $W^+$ and $W^-$ respectively.  
In this note we focus on a single-differential distribution and study the azimuthal distribution of the decay products of one of the two $W$'s. 
 We remain inclusive about the other $W$, which can then  be thought of as a state of well defined helicity. The interference term between the transverse-transverse amplitudes reads \cite{Panico:2017frx}
\begin{eqnarray}\label{intWg}
&&I^{WW}\hspace{-2pt}\propto {\mathcal{A}}_{++}^{\textrm{\tiny{BSM}}}\hspace{-2pt}
\left[\hspace{-1pt}{\mathcal{A}}_{-+}^{\textrm{\tiny{SM}}}\hspace{-4pt}+\hspace{-2pt}{\mathcal{A}}_{+-}^{\textrm{\tiny{SM}}}\hspace{-2pt}
\right]\hspace{-2pt}\cos{2\varphi}\,,
\end{eqnarray}
where the  angle $\varphi$  is measured  making reference to the outgoing fermion of positive helicity (\(\varphi = \varphi^+\) or \(\varphi^-\), depending on whether \(W^+\) or \(W^-\) is chosen). Consistently, the interference \eq{intWg}  vanishes when integrated over, reproducing  non-interference results.

From the form of these parton-level amplitudes Eqs.~(\ref{bsmTT},\ref{ampssm}), we deduce the following important aspects. \emph{i)}   The backward region $\cos\Theta\approx -1$ is not favourable, as  SM and BSM both vanish.
\emph{ii)}  The BSM amplitude has its maximum in the central region  $\cos\Theta\approx 0$, but the SM switches from being dominated by the $+-$ to being dominated by $-+$, which has opposite sign: a $\cos\Theta$-inclusive analysis has therefore the potential to partially cancel the interference term.
 \emph{iii)} The forward region $\cos\Theta\approx +1$  has a large SM occupations because of the $t$-channel neutrino pole, while BSM vanishes; interference is in fact finite and the signal over square-root of background increases rapidly $\sim\Theta^{3/2}$ as we approach the central region.

The angle $\varphi$  is measured  with respect to  the outgoing fermion of positive helicity, which is indistinguishable from a fermion of opposite helicity when the $W$ decays hadronically, implying an ambiguity\footnote{Jet substructure techniques for tagging $W$-boson decays to charm/strange quarks or identifying charge could offer a handle on this.}
\begin{equation}\label{eq:amb}
\varphi\leftrightarrow \varphi+\pi.
\end{equation}
Fortunately, distributions of the form Eq.~(\ref{intWg}) are insensitive to this ambiguity.
The fully hadronic channel makes it also impossible to distinguish $W^+$ and the $W^-$, leading to an extra
\begin{equation}\label{ambbigtheta}
\Theta\to\Theta+\pi\,
\end{equation}
ambiguity. This is equivalent to $\cos\Theta\to-\cos\Theta$ under which, see point \emph{ii)} above, the interference term is odd close to $\cos \Theta\approx 0$. Therefore, in the very central region, the interference cancels in the fully hadronic channel.

For leptonically decaying $W$-bosons in the semileptonic channel $\nu l^+\bar q q $, an angle could be defined  with reference to the charged lepton, and this could be related to $\varphi$. Yet, a decay plane can be established only with knowledge of the neutrino momentum, which can be reconstructed from missing  energy and the $W$-decay kinematics, up to a twofold ambiguity, where the two solutions differ by their longitudinal momentum.
 At lepton colliders---despite the fixed total center-of-mass energy---initial state radiation (ISR) and beamstrahlung
  smear the  initial energy spectrum, making  the leptonic-collider case similar in practice 
   to the LHC: the initial momentum on the beam-pipe direction is not known with sufficient precision.
This implies  an approximate ambiguity 
\begin{equation}\label{eq:amb2}
\varphi \leftrightarrow \pi- \varphi
\end{equation}
However, this doesn't affect the distribution \eq{intWg} (in constrast, CP odd effects from $W^{a\, \nu}_{\mu}W^{b}_{\nu\rho}\widetilde W^{c\, \rho\mu}$ give distributions $\propto\sin 2\varphi$, which are odd under \eq{eq:amb2} and therefore erased by this ambiguity). 

\paragraph{Selection Cuts.}
This understanding guides us in designing the experimental analysis. 
We divide $\varphi\in[-\pi,\pi]$ in 10 bins. For the semileptonic channel we can choose $\varphi$ to be the angle defined by the leptonically decaying  $W$; the results do not change if we choose the hadronically decaying one.
We focus on the channel $W^-\to \mu^-\bar\nu$ and multiply by a factor 4 the luminosity to account for $l=\mu^+$ and $l=e^\pm$.\footnote{In principle the analysis with $l=e^\pm$ is complicated because of a $t$-channel diagram not present in the muon channel; however, we have verrified that, due to its different kinematics, this effect can be efficiently singled out by using polar angle and lepton transverse mass cuts.} Finally, we separate the polar angle as $\cos\Theta\in [-1,-0.5,0,0.5,1]$ for the semileptonic channel.
For the fully hadronic channel, following point \emph{ii)} and the ambiguity \eq{ambbigtheta}, we instead single out the very central region where interference is suppressed, $\cos\Theta\in [-1,-0.5,-0.2,0,0.2,0.5,1]$. Furthermore, for this channel, we define $\varphi$ by selecting one of the two fermions randomly and define $\Theta$ by selecting randomly one of the two $W$'s, reflecting the ambiguity \eq{eq:amb} and \eq{ambbigtheta}.

The broad  energy-spectrum created by ISR and beamstrahlung, convoluted with the fact that the cross section increases towards smaller energy,  implies that a good fraction of events has energy  smaller than the nominal collider energy $E^{nom}$.
This complicates the azimuthal-angle analysis since, at small energies, mixed helicity amplitudes $\pm0$, $0\pm$ appear in the SM and interfere with $\pm\mp$, generating  non-trivial azimuthal distributions that make it difficult to recognise distribution of the form Eq.~(\ref{intWg}). For this reason, we perform selection cuts on the energy of the events,  as shown in the last column of table~\ref{tab:ILC/CLICbaseline}.
\begin{table}[h]
\begin{center}
\begin{tabular}{c|c|c|c|}
\cline{2-4}
&{\bf Energy (GeV)} & \multicolumn{1}{ c| }{\textbf{Luminosity (ab $^{\mathbf{-1}}$)}} & {\bf Analysis Cut}\\  \hline
\multicolumn{1}{ |c| }{\multirow{3}{*}{\textbf{CLIC}}} & {3000}  & $4_L+1 _R$ & $\sqrt{s}>2600$ \\
 \cline{2-4}
\multicolumn{1}{ |c| }{} &  {1500}   &  $2_L+0.5 _R$  & $\sqrt{s}>1300$\\
 \cline{2-4}
\multicolumn{1}{ |c| }{} &  {380}   &  $0.5_L+0.5 _R$   & $\sqrt{s}>330$\\
 \cline{1-4}
 \multicolumn{1}{ |c| }{\multirow{2}{*}{\textbf{ILC}}} & {500}  & {4}  & $\sqrt{s}>400$ \\
 \cline{2-4}
\multicolumn{1}{ |c| }{} &  {250}   & {2} & $\sqrt{s}>200$\\
 \cline{1-4}
\end{tabular}
\end{center}
\caption{\it CLIC and ILC run scenarios assumed in this analysis. Subscript $L(R)$ denote runs with electron beams polarized at 80\% Left (Right). The last column denotes additional cuts imposed on our analysis to select genuinely high-energy events with reduced impact from beamstrahlung and ISR.}\label{tab:ILC/CLICbaseline}
\end{table}


\begin{table}[h]
\centering
\begin{center}
$\mathbf{c_{3W}\times 10^{2}}$\\
\makebox[\textwidth]{\begin{tabular}{|c|c|c|c|c|c|}
\hline 
\multicolumn{1}{ |c|  }{\multirow{2}{*}{$\mathbf{e^+e^-\rightarrow W^+W^-}$}} &\multicolumn{1}{ |c|  }{\multirow{2}{*}{$\mathbf{\sqrt{s}}$}} &\multicolumn{2}{ c| }{\textbf{Fully Hadronic }} &\multicolumn{2}{ c| }{\textbf{Semileptonic }} \\
  &  &  Inclusive $\varphi$ & Exclusive $\varphi$& Inclusive $\varphi$ & Exclusive $\varphi$  \\ 
\hline
\multicolumn{1}{ |c  }{\multirow{3}{*}{\textbf{CLIC}$\mathbf{(3\%)}$}} &
\multicolumn{1}{ |c| }{$3$ TeV}  & [-3.40, 3.54] & [-1.99, 2.01] & [-2.79, 3.22] & [-1.13, 1.16] \\ \cline{3-6}
\multicolumn{1}{ |c  }{}    &   \multicolumn{1}{ |c| }{$1.5$ TeV}  & [-11.8, 13.0] & [-6.52, 6.90] &  [-10.1, 11.0] & [-3.38, 3.42]\\ \cline{3-6}
\multicolumn{1}{ |c  }{}   &    \multicolumn{1}{ |c| }{$380$ GeV}  & [-82, 173] & [-37.4, 43.9] & [-61.5, 89.1] & [-23.8, 25.2] \\ \cline{1-6}
\multicolumn{1}{ |c  }{\multirow{3}{*}{\textbf{CLIC}($\mathbf{1\%}$)}} &
\multicolumn{1}{ |c| }{$3$ TeV}  & [-3.14, 3.26] & [-1.81, 1.83] & [-2.64, 3.07] & [-1.07, 1.01]\\ \cline{3-6}
\multicolumn{1}{ |c  }{}     &     \multicolumn{1}{ |c| }{$1.5$ TeV}& [-10.3, 11.5] & [-5.67, 5.92]  & [-9.06, 10.1] & [-3.12, 3.16] \\ \cline{3-6}
\multicolumn{1}{ |c  }{}     &     \multicolumn{1}{ |c| }{$380$ GeV} & [-56.8, 80.8] & [-30.4, 34.4] & [-42.7, 53.6] & [20.1, 21.0]\\ \hline
\multicolumn{1}{ |c  }{\multirow{2}{*}{\textbf{ILC}($\mathbf{3\%}$)}} &
\multicolumn{1}{ |c| }{$500$ GeV} & [-63, 124] & [-15, 15] \\ \cline{3-4}
 \multicolumn{1}{ |c  }{}      &    \multicolumn{1}{ |c| }{$250$ GeV} & [-208, 301] & [-54, 55] \\ \cline{1-4}
\multicolumn{1}{ |c  }{\multirow{2}{*}{\textbf{ILC}($\mathbf{1\%}$)}} &
\multicolumn{1}{ |c| }{$500$ GeV} & [-42, 72] & [-10, 10] \\ \cline{3-4}
\multicolumn{1}{ |c  }{}      &     \multicolumn{1}{ |c| }{$250$ GeV} & [-128, 157] & [-34, 35] \\ \cline{1-4}
\end{tabular} }
\quad
\caption{ \it Sensitivity on $c_{3W}\times10^2$ at 68\% C.L. at CLIC and ILC for hadronic and semileptonic channels and  different systematic uncertainties (in parentheses), run conditions and analysis strategy; the first (second) columns denotes inclusive (exclusive) analysis in the azimuthal angle $\varphi$.}
\label{tab:Transverseresults}
\end{center}
\end{table}

\begin{figure}[h]
\centering
\includegraphics[width=0.48\textwidth]{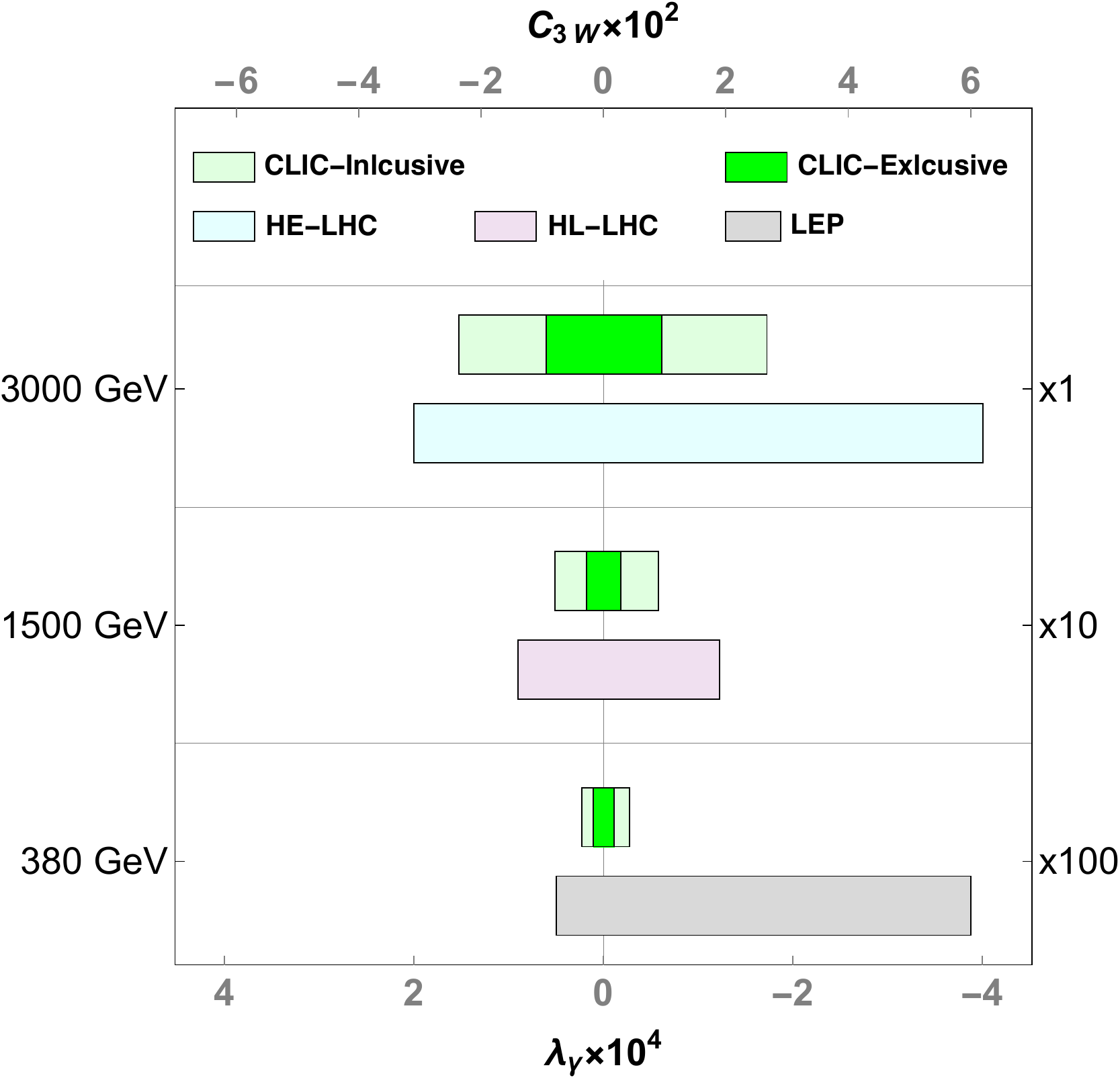}
\includegraphics[width=0.48\textwidth]{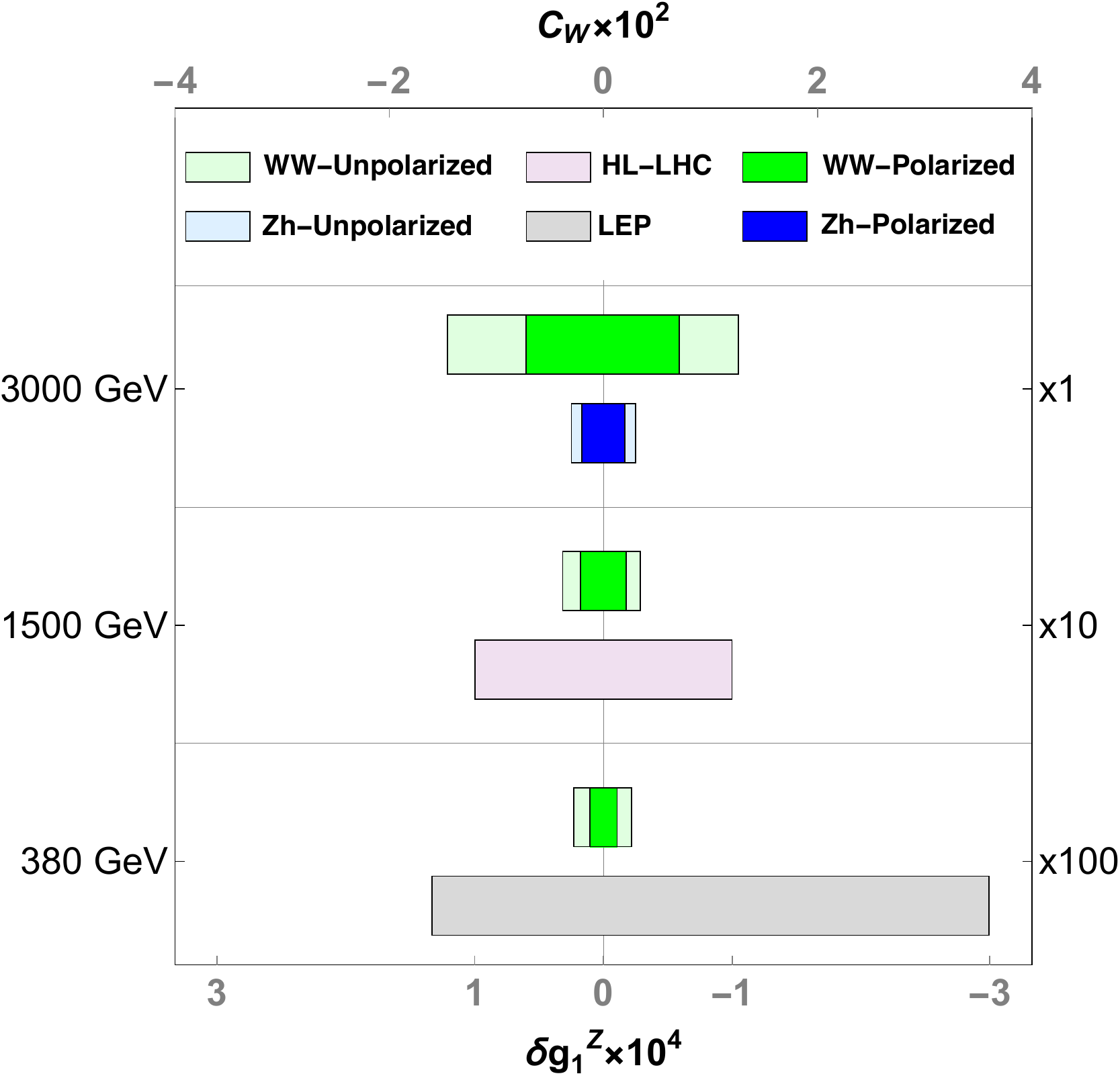}
\caption{\it Bounds on the Wilson coefficient $c_{3W}$ (left) $c_W$ (right) or equivalently in the BSM parameter $\lambda_\gamma$ and $\delta g_1^Z$. The systematics is set to 1\%; the LEP and LHC results are taken from the references in the text.}
\label{fig:finplot}
\end{figure}

\paragraph{Analysis.} We generate events using  {\sc Whizard}~\cite{Kilian:2007gr} according to the energy/luminosity scenarios in table~\ref{tab:ILC/CLICbaseline},  from \cite{deBlas:2018mhx}, and take into account  detector effects  by  smearing the energy of jets on a normal distribution with a 4\% resolution.  We also require the polar angle  for jets and leptons to fulfil  $10^{\circ}<\Theta_{l,j}<170^{\circ}$ and $-0.95<\cos\Theta<0.95$ for the reconstructed $W$s to avoid the forward region; in addition we impose $E>10$~GeV  for all particles and $M_{jj}>10$ GeV, to avoid events from virtual photons in the fully hadronic case.
We then compare the different energy stages at CLIC and ILC, and scenarios with optimistic 1\% and pessimistic 3\% systematic uncertainty $\delta_{syst}$ in all bins, where we also include a  50\%  signal acceptance.
The results are summarized in table \ref{tab:Transverseresults}, and in the left panel of figure~\ref{fig:finplot}. We compare analyses with binned $\varphi$ against unbinned analyses where all other cuts are identical.
As expected, in both the  channels, the analysis binned in the azimuthal angle, which takes into account the interference effects, gives additional sensitivity; moreover, we verify that the sensitivity is dominated by the central bins in $\Theta$, as expected from our discussion of points \emph{i)} and \emph{iii)} above. Our results also confirm the rough $E^2$ sensitivity improvement at high-energy. The results for the fully hadronic analysis and the semileptonic one are almost identical; this is due to the larger  luminosity of the former  being  compensated by the ambiguity in the polar distribution, as discussed in point \emph{ii)}.

Diboson processes were already scrutinised at LEP II \cite{LEP:2003aa}, where aTGCs were constrained as $\lambda_\gamma\in[-0.04,0.005]$. Similar searches are performed at the LHC, and it is believed that, by the end of the HL-LHC program (3 ab$^{-1}$), the sensitivity to these effects will have improved by more than an order of magnitude, depending on the final state \cite{Panico:2017frx,Azatov:2017kzw,Azatov:2019xxn}; further improvements are forecasted for the putative 27 TeV (15 ab$^{-1}$) HE-LHC extension, which we include for comparison. At 68\% C.L,
\begin{align}
\text{HL-LHC:}  \quad &  \lambda_\gamma\in[-2.1, 1.2]\times10^{-3} \,\,  (WZ) & \lambda_\gamma\in [-1.2, 0.9]\times10^{-3} \,\,(W\gamma)\nn\\
\text{HE-LHC:} \quad&  \lambda_\gamma\in[-0.7, 0.4]\times10^{-3}  \,\, (WZ), &\lambda_\gamma\in [-0.4, 0.2]\times10^{-3}\,\,(W\gamma)\,,
\end{align}
where the results from $WZ$ final states stem from Refs.~\cite{Azatov:2017kzw,Azatov:2019xxn}, while the $W\gamma$ results from Ref.~\cite{Panico:2017frx}.
These  are sumarized in fig.~\ref{fig:finplot}; notice that, while the 1.5 TeV-run is competitive with HE-LHC, the 3 TeV one is about a factor three better.

\subsection{Longitudinal  $W^+W^-$ and $Zh$ Final States}\label{sec:anlong}
New physics in the Higgs sector manifests itself both in processes with longitudinal polarizations $W^+_LW^-_L$ and $Z_Lh$ associated production. At high-energy only one dimension-6 effect survives for each process and each helicity configuration  \cite{Jacob:1959at,Franceschini:2017xkh}; we will therefore focus our analysis on the operators $\op_W$ and $\op_B$.

In the high-energy tree-level regime, the SM amplitudes are
\begin{gather} \label{ampssmlogSM}
({\cal A}_{\textrm{\tiny{SM}}})^{00}_{\text{L}}\approx -\frac{1}{4}(g^2+{g^\prime}^2) \sin\Theta,\quad\quad
({\cal A}_{\textrm{\tiny{SM}}})^{Zh}_{\text{L}} \approx -\frac{1}{4}(g^2-{g^\prime}^2) \sin\Theta\,,\\
({\cal A}_{\textrm{\tiny{SM}}})^{00}_{\text{R}}\approx-({\cal A}^{Zh}_{\textrm{\tiny{SM}}})_{\text{R}}\approx -\frac{1}{2}{g^\prime}^2 \sin\Theta\,, \nn
\end{gather}
where $\Theta$ corresponds to the angle between the incoming electron and the outgoing $W^-$ or~$Z$. BSM effects induce the following  energy-growing effects relative to the SM, $\Delta\cal A_{\textrm{\tiny{BSM}}} = {\cal A}_{\textrm{\tiny{BSM}}}/\cal A_{\textrm{\tiny{SM}}}$
\begin{gather} \label{ampssmlog}
(\Delta{\cal A}_{\textrm{\tiny{SM}}})^{00}_{\text{L}}\approx(\textit{c}_w^2c_W + \textit{s}_w^2c_B)\frac{s}{\Lambda^2}\quad\quad
(\Delta{\cal A}_{\textrm{\tiny{SM}}})^{Zh}_{\text{L}} \approx\dfrac{1}{\textit{c}_w^2-\textit{s}_w^2} (\textit{c}_w^2c_W - \textit{s}_w^2c_B )\frac{s}{\Lambda^2} \nn\\
(\Delta{\cal A}_{\textrm{\tiny{SM}}})^{00}_{\text{R}}\approx (\Delta{\cal A}_{\textrm{\tiny{SM}}})^{Zh}_{\text{R}} \approx c_B \frac{s}{\Lambda^2}\quad\quad \,,
\end{gather}
with $s_w$ and $c_w$ the sine and cosine of the weak mixing angle.
Here we treat $\op_W$ and $\op_B$ as independent (i.e. the $S$-parameter $\sim c_W+c_B$ is free to vary, see also Ref.~\cite{Zhang:2016zsp}). 
Notice that  modifications of the input parameters  do not induce energy-growth, as they must be proportional to the SM cross-sections.


\paragraph{Selection Cuts and Analysis.} For longitudinal $W^+W^-$ final states, the analysis is complicated by the fact that  the SM amplitude ${\cal A}^{00}_{SM}$ is much smaller than the transverse ones \eq{ampssm}. This is due in part to the fact that there are more transverse helicity configurations,  in part to the sum over  $SU(2)_L$ group factors that appear enhanced in the transverse amplitude, and in part to the forward enhancement from the $t$-channel singularity of the transverse amplitude. 
In inclusive (longitudinal + transverse) measurements, this large transverse fraction effectively acts as background at high-energy. However, at small energy the transverse/longitudinal interference is enhanced by finite-mass effects. So, in practice, the interference, though present, is suppressed. 

The $t$-channel diagram entering the transverse amplitude---in which a neutrino is exchanged---only involves left-handed electrons (and right-handed positrons). For this reason, a polarization of  the initial beam that preserves only  right-handed electrons (and left-handed positrons) would suppress the transverse contribution and make the analysis of the longitudinal components more efficient. A strategy along these lines is mentioned in Ref.~\cite{Battaglia:2004mw}; strangely enough, however, it is applied indiscriminately also to searches for $\lambda_\gamma$ (equivalently, \(c_{3W}\)), where it doesn't improve the sensitivity, since $({\cal A}^{+-}_{\textrm{\tiny{SM}}})_{\text{R}}\approx 0$. Here we combine the  benefits of partially polarized beams together with differential distributions in the polar angle, an important improvement given that CLIC polarization setups are not 100\% and  transversely polarized vectors still provide an overall important background.

We perform the $W^-_L W^+_L$  analysis using the same cuts as for the  above $W^-_T W^+_T$ study, except from the binning in $\varphi$ that plays no major role here and is ignored. For technical reasons related with the Whizard implementation of BSM effects, we only simulate the effects of $\op_W$ and discuss how to recover the results for $\op_B$ in section~\ref{sec:interp}.\footnote{We use an implementation of dimension-6 operators based on aTGCs which includes a single effect \(\delta g_1^Z\). As mentioned earlier and discussed further in sec.~\ref{sec:interp}, the high-energy limit only receives a contribution from a certain linear combination of \(\mathcal{O}_W\) and \(\mathcal{O}_B\).}
The results are summarized in table \ref{tab:Longitudinalsresults} and in the right panel of figure \ref{fig:finplot}. Since here the azimuthal distributions are not highlighted, the more luminous fully hadronic channels give the tightest constraint as expected.

Searches fo these effects at LEP \cite{LEP:2003aa}, gave $c_W\in [-1.6, 3.6]$, while forecasts for the HL-LHC \cite{Franceschini:2017xkh} imply a reach $c_W\in [-0.12, 0.12]$;  we show these in figure \ref{fig:finplot} for comparison\footnote{For completeness, we also mention that the previous CLIC-Report forecasts \cite{Battaglia:2004mw} are slightly more pessimistic but still quite in agreement with our results.}.

\begin{table}[h]
\centering
\begin{center}
$\mathbf{c_{W}\times 10^{2}}$\\
\makebox[\textwidth]{\begin{tabular}{|c|c|c|c|c|c|}
\hline 
\multicolumn{1}{ |c|  }{\multirow{2}{*}{$\mathbf{e^+e^-\rightarrow W^+W^-}$}} &\multicolumn{1}{ |c|  }{\multirow{2}{*}{$\mathbf{\sqrt{s}}$}} &\multicolumn{2}{ c| }{\textbf{Fully Hadronic }} &\multicolumn{2}{ c| }{\textbf{Semileptonic }} \\
&  &  Unpolarized & Polarized & Unpolarized & Polarized  \\ \hline
\multicolumn{1}{ |c  }{\multirow{3}{*}{\textbf{CLIC}$\mathbf{(3\%)}$} } &
\multicolumn{1}{ |c| }{$3$ TeV}  & [-1.98, 1.62] & [-0.96, 0.94] & [-3.43, 2.50] & [-1.61, 1.53] \\ \cline{3-6}
\multicolumn{1}{ |c  }{}    &        \multicolumn{1}{ |c| }{$1.5$ TeV}  & [-5.95, 5.08] & [-3.17, 3.15] &  [-8.80, 7.04] & [-4.30, 4.11]\\ \cline{3-6}
\multicolumn{1}{ |c  }{}   &         \multicolumn{1}{ |c| }{$380$ GeV}  & [-62.6, 56.4] & [-23.4, 24.1] & [-56.8, 51.1] & [-22.8, 23.2] \\ \cline{1-6}
\multicolumn{1}{ |c  }{\multirow{3}{*}{\textbf{CLIC}$\mathbf{(1\%)}$} } &
\multicolumn{1}{ |c| }{$3$ TeV}  & [-1.71, 1.44] & [-0.78, 0.77] & [-3.64, 2.60] & [-1.52, 1.45]\\ \cline{3-6}
\multicolumn{1}{ |c  }{}     &    \multicolumn{1}{ |c| }{$1.5$ TeV}& [-4.69, 4.14] & [-2.57, 2.53]  & [-8.09, 6.60] & [-3.91, 3.80] \\ \cline{3-6}
\multicolumn{1}{ |c  }{}     &     \multicolumn{1}{ |c| }{$380$ GeV} & [-38.0, 35.6] & [-16.7, 17.0] & [-41.2, 38.1] & [-19.0, 19.10]\\ \hline
\multicolumn{1}{ |c  }{\multirow{2}{*}{\textbf{ILC}$\mathbf{(3\%)}$}} &
\multicolumn{1}{ |c| }{$500$ GeV} & [-35, 32] & \multicolumn{1}{ |c  }{} \\ \cline{3-3}
 \multicolumn{1}{ |c  }{}      &   \multicolumn{1}{ |c| }{$250$ GeV} & [-157, 142] & \multicolumn{1}{ |c  }{} \\ \cline{1-3}
\multicolumn{1}{ |c  }{\multirow{2}{*}{\textbf{ILC}$\mathbf{(1\%)}$} } &
\multicolumn{1}{ |c| }{$500$ GeV} & [-20, 19] & \multicolumn{1}{ |c  }{}  \\ \cline{3-3}
\multicolumn{1}{ |c  }{}      &    \multicolumn{1}{ |c| }{$250$ GeV} &  [-90, 84] & \multicolumn{1}{ |c  }{}  \\ \cline{1-3}
\end{tabular} }
\quad
\caption{ \it One sigma sensitivity on $\bm{c_{W}}$ from longitudinal $WW$ final states, at CLIC and ILC for both the hadronic and semileptonic channels and for different systematic uncertainties, and different polarization setups as defined in table \ref{tab:ILC/CLICbaseline}.}
\label{tab:Longitudinalsresults}
\end{center}
\end{table}

\paragraph{Associated Production.} As argued above, high-energy $Zh$ production is sensitive to similar effects as those entering $WW$ processes. This channel includes however  no relevant background: it is a rather simple process in which SM and BSM have identical angular distributions and interfere maximally, so that a simple counting analysis suffices.\footnote{More sophisticated angular-distribution analyses grant access to other BSM classes of effects in which the Higgs couples to transversely polarized gauge bosons~\cite{Craig:2015wwr,Beneke:2014sba}.} 

We use {\sc MadGraph} \cite{Alwall:2014hca} to simulate ILC and CLIC, focussing on the highest energy stages, which offer the best sensitivity. We assume complete reconstruction of the final states. Additionally, we take into account acceptance, ISR and brehmstrahlung (not included in Madgraph) as an effective 50\% reduction with respect to the nominal luminosity: this reflects the efficiency of the cuts in the third column of table~\ref{tab:ILC/CLICbaseline}. 
The resulting reach is reported in table \ref{tab:Zhresults}.

\begin{table}[ht]
\begin{center}
\begin{tabular}{|c|c|c|c|}
\multicolumn{4}{ c }{$\mathbf{c_W\times 10^{-3}}$}\\
\hline 
\multirow{1}{*}{$\mathbf{e^+e^-\rightarrow Zh}$} & \multicolumn{1}{ |c|  }{\multirow{1}{*}{$\mathbf{\sqrt{s}}$}} & Unpolarized & Polarized \\ \hline
\multicolumn{1}{ |c|  }{\multirow{1}{*}{ \textbf{CLIC}(\textbf{3\%})}}  & \multicolumn{1}{ |c|  }{$3$ TeV} & $[-4.6, 4.7]$ & $[-2.8, 2.9]$ \\ \hline
\multicolumn{1}{ |c|  }{\multirow{1}{*}{\textbf{CLIC}(\textbf{1\%})}} &  \multicolumn{1}{ |c|  }{$3$ TeV} & $[-2.6, 2.6]$ & $[-1.7, 1.7]$ \\ \hline
\multicolumn{1}{ |c|  }{\multirow{1}{*}{\textbf{ILC}(\textbf{3\%})}} & \multicolumn{1}{ |c|  }{$0.5$ TeV} & $[-20, 21]$ & \multicolumn{1}{ |c }{} \\ \cline{1-3}
\multicolumn{1}{ |c|  }{\multirow{1}{*}{\textbf{ILC}(\textbf{1\%})}} & \multicolumn{1}{ |c|  }{$0.5$ TeV} & $[-7, 7.1]$ &  \multicolumn{1}{ |c }{}\\ \cline{1-3}
\end{tabular}
\quad
\caption{\it One sigma sensitivity on $\bm{c_{W}}$ at CLIC and ILC in $Zh$-production.
The comparison is between the results from highest energy runs, see the text for more details on the analysis.}
\label{tab:Zhresults}
\end{center}
\end{table}

\section{Interpretation}\label{sec:interp}

\subsection{BSM in the Transverse Sector}
In section \ref{sec:antrans} we have seen that dedicated analyses of $WW$ at future linear colliders potentially provide an unprecedented precision on BSM interactions stemming from modifications of the transverse sector, a precision of the order of $\sim 10^{-4}$ when expressed in terms of anomalous TGCs. The relevant question, however, is whether this precision would provide us with a better understanding of putative BSM dynamics. To answer this question, in this section we translate our results into BSM reach, for the weak and strong coupling scenarios discussed in section~\ref{sec:BSM} (\eq{eq:C2W3W} and \eq{remediosscaling}).

First of all, it is important to accept that new physics that interacts with the transverse polarizations of vectors,  generically modifies their self interactions (triple vertices), as well as  the $W/Z$ propagators,  as discussed in section~\ref{sec:BSM}.  The latter induces energy-growing effects in Drell-Yann (DY) processes, which offer a fantastic probe for new physics, as discussed in Refs.~\cite{Barbieri:2004qk,Farina:2016rws,Alioli:2017jdo,deBlas:2018mhx}. 

\begin{figure}[t]
\centering
\includegraphics[width=0.7\textwidth]{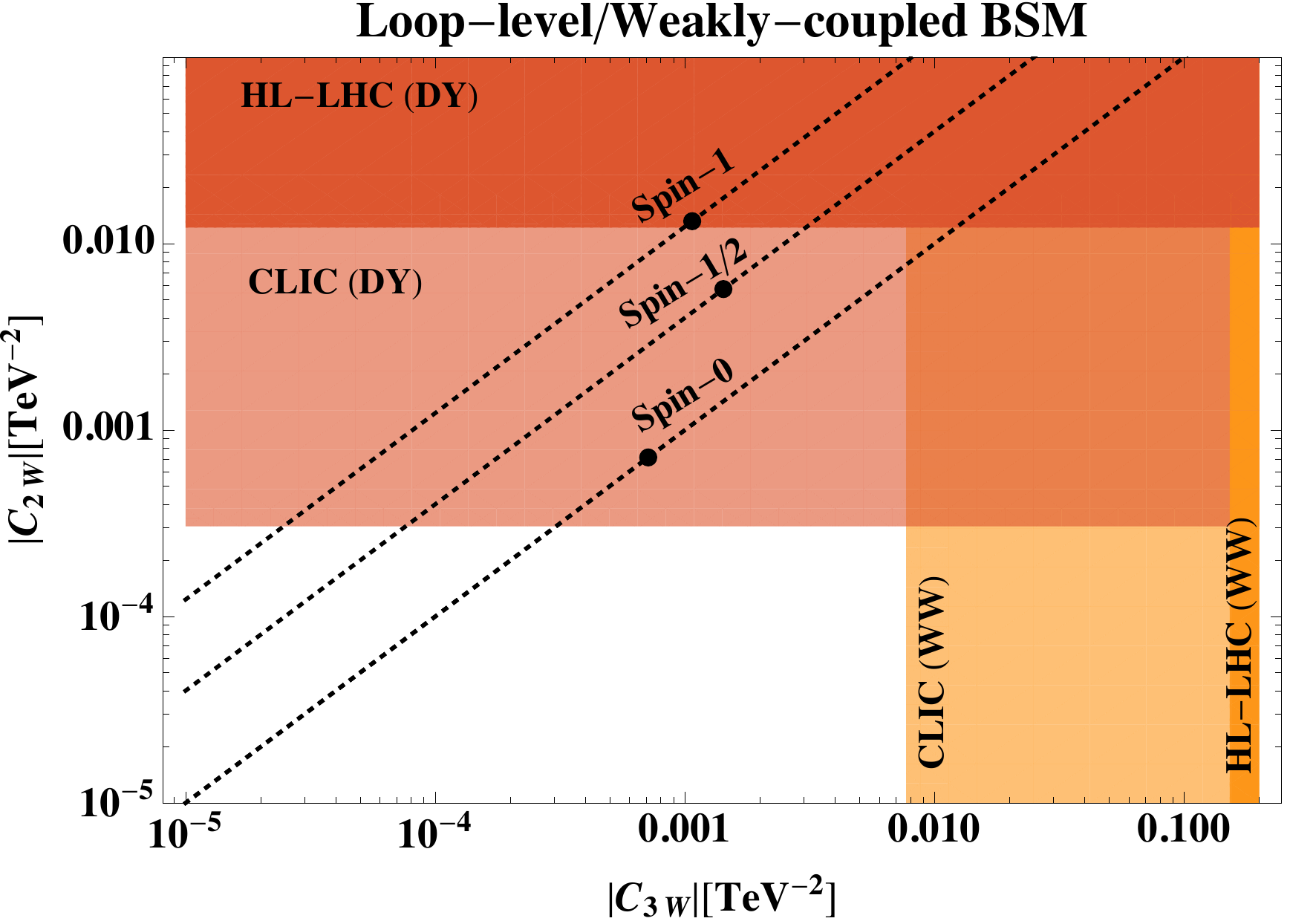}
\caption{\it Ultimate reach of HL-LHC and CLIC on the BSM-coefficients $C_{2W/3W}$ for dibosons and Drell-Yan processes. The dashed lines indicate the theoretical prediction in weakly coupled models for different spins of the heavy BSM states (spin-0 denotes complex scalars), while the solid dots on them highlight the value for a particle with mass just beyond the CLIC 3 TeV reach, and in a gauge representation with $\mu(R)=2$. }
\label{fig:transverse}
\end{figure}

\begin{figure}[h]
\centering
\includegraphics[width=0.7\textwidth]{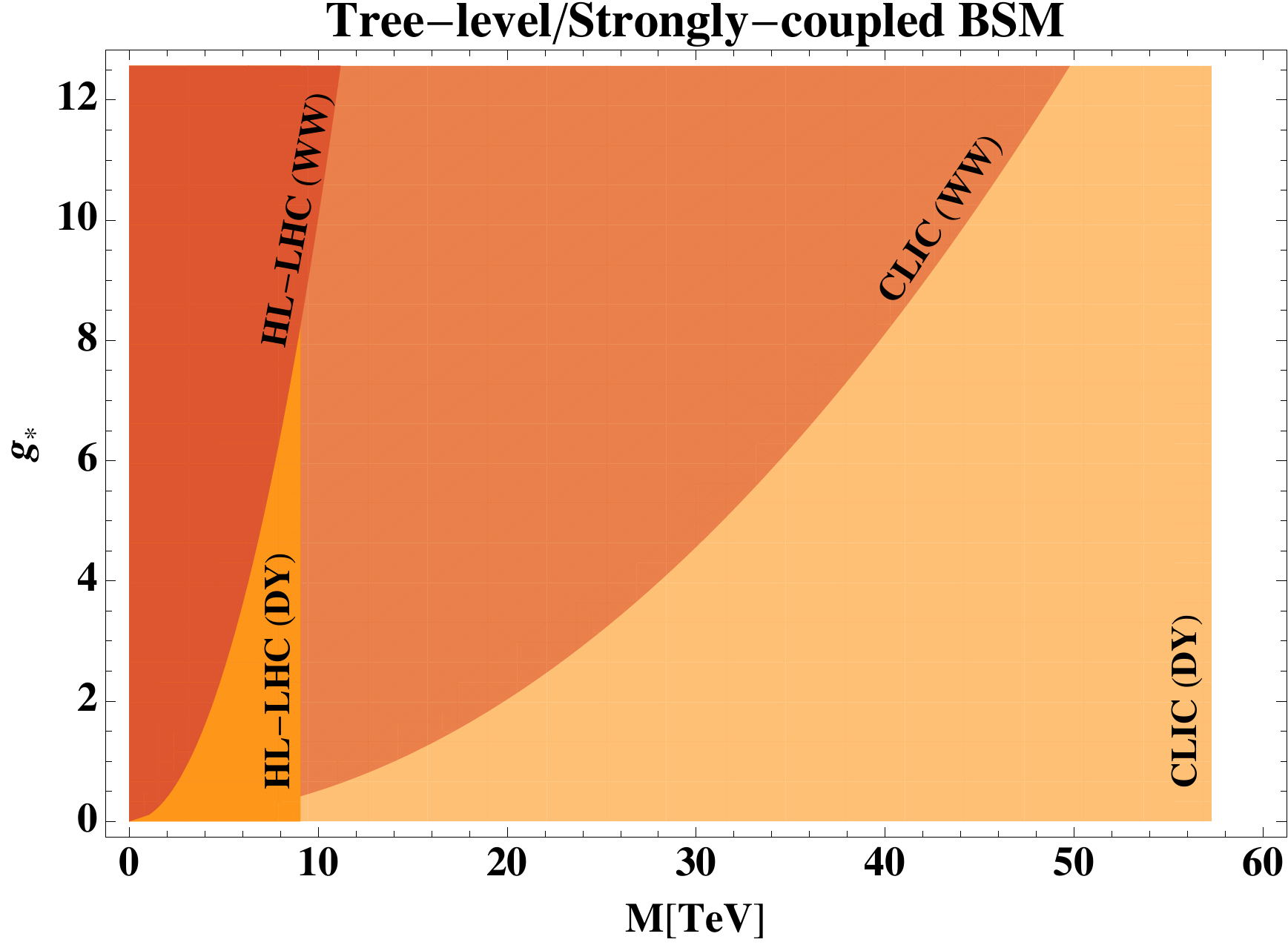}
\caption{\it Comparison of the reach of Drell-Yan and Dibosons at HL-LHC and CLIC, on the strongly-coupled BSM scenarios of Ref.~\cite{Liu:2016idz}. Here $g_*$ denotes the coupling and $M$ the mass scale associated with the new BSM sector.}
\label{fig:transverse2}
\end{figure}

In figs~\ref{fig:transverse} and \ref{fig:transverse2} we compare the reach from DY processes with the reach of our $WW$ analysis in different models, where we base CLIC results  on the final 3 TeV stage only. DY results are taken from Refs.~\cite{Farina:2016rws,deBlas:2018mhx} for HL-LHC and CLIC respectively, while diboson HL-LHC results are from Ref.~\cite{Panico:2017frx}.
 
\paragraph{Weak Coupling.} Figure~\ref{fig:transverse} focusses on weakly coupled models, as discussed around  \eq{eq:C2W3W}. The dotted lines denote the theoretical predictions in situations where the EFT effects are dominantly produced by particles of a given spin 0, $1/2$ or 1, according to  table~\ref{eq:a2W3W}. Generally, DY processes are likely to  first discover deviations from the SM. In fact, for particles of spin-1 or $1/2$, any effect that is visible in $WW$ processes at the last CLIC stage, should have already been observed at the HL-LHC in DY. For particles of spin-0 instead, it is possible to that both will show up at CLIC only.

Nevertheless, if experiments find evidence for \(C_{2W} \ne 0\) and \(C_{3W} \ne 0\), then the size of these coefficients clues us into the mass scale (\(C_{2W,3W} = c_{2W,3W}/\Lambda^2\)) while their ratio clues us into the spin, thereby providing valuable information on the most important properties---the kinematic information---about the BSM physics.\footnote{This conclusion also holds for the exceptional case where the BSM physics contains neutral, triplet vectors that generate \(\mathcal{O}_{2W}\) at tree-level. In this case \(C_{2W} \approx C_{2W}^{\text{tree}} = \tilde{g}^2/M^2\) with \(\tilde{g}\) a coupling constant of the UV physics~\cite{Henning:2014wua}. In this case, for order-one coupling constants, \(C_{2W}/C_{3W} \sim (4\pi)^2\) is large and easily distinguished.}

Relations such as this one---which depend on kinematic features of the BSM sector rather than its details---are very valuable.
Recently, the authors of~\cite{Quevillon:2018mfl} obtained results analogous to those in eq.~\eqref{eq:C2W3W} for a variety of dimension-6 and 8-operators that contribute to the 2-, 3-, and 4-point interactions of EW gauge bosons. Similar to our discussion in sec.~\ref{sec:BSM}, one notes that various ratios of these quantities depend only on the spin of the underlying particle; it would be interesting to understand the general principle at work here.

The black dots in figure \ref{fig:transverse} denote the predictions from models with $\mu(R)=2$ in \eq{eq:C2W3W} and a mass at the very edge of on-shell discovery at CLIC. Interestingly, while DY can consistently probe these theories, $WW$ processes are somewhat penalised. This is due partially to the smaller cross sections in $WW$, partially to the difficulties of resurrecting the interference, and partially to a numerical accident that exhibits enhanced DY effects over $WW$ ones, see table~\ref{eq:a2W3W}. Larger $\mu(R)$, instead, imply larger effects, and can be consistently tested also in $WW$; for instance a Dirac Fermion in an $SU(2)_L$ representation with $\mu(R)\sim 10$ and mass above CLIC threshold is visible also in $WW$ processes and consistently described by the $\op_{3W}$ EFT.

\paragraph{Strong Coupling.}

In figure  \ref{fig:transverse2}, we focus on the models with multipolar interactions of Ref.~\cite{Liu:2016idz}, which assume new strongly coupled dynamics at the scale $M$, as captured by \eq{remediosscaling}. As the coupling strength $g_*$ increases, the $WW$ channel becomes more relevant compared to DY. However, even at maximal strong coupling, the $WW$ analysis does not seem competitive. Nevertheless we recall that the estimates of \eq{remediosscaling} are parametric as the scenarios of Ref.~\cite{Liu:2016idz} do not have a weakly coupled and calculable analog. For this reason, $WW$ processes  still provide valuable information on these types of models and might even be more sensitive than DY.

\subsection{BSM in the Higgs sector}

The comparison between the reach of high-energy $Zh$ associated production and processes with longitudinally polarized $WW$ shows that, in terms of a single parameter, chosen to be the coefficient of $\op_W$ in section \ref{sec:reach}, $Zh$ performs much better.
{In particular, the CLIC sensitivity translates roughly to scales of order $\sim25$ TeV in terms of the mass of heavy vector triples mentioned in \cite{Franceschini:2017xkh}}.

Nevertheless, as shown in Eq.~(\ref{ampssmlog}), $Zh$ and $WW$ are sensitive to different combinations of $c_W$ and $c_B$, and their combined information allows us to reach this 2-dimensional parameter space. Notice that the combination $\hat s=(c_W+c_B)m_W^2/\Lambda^2$ corresponds to the $S$ parameter \cite{Peskin:1991sw,Barbieri:2004qk}, which was measured with per-mille precision at LEP-I and is a major motivation for the construction of circular $e^+e^-$ colliders operating for extended periods of time on the $Z$-pole resonance. 

We focus on the high-energy CLIC run at 3 TeV, where the expressions for the amplitudes in the massless limit approximate well the SM and BSM  amplitudes, and translate the results of section \ref{sec:anlong} (performed in terms of a unique parameter), into a combined analysis for $c_W$ and $c_B$ using \eq{ampssmlogSM} and \eq{ampssmlog}.\footnote{We approximate the left/right polarised beam, which contains 90\% of left-handed/right-handed electrons, with a 100\% polarised beam; additionally, for the right-handed we focus on the central bins of polar angle, where this approximation works particularly well, as the number of transverse is sensibly lower.} As motivated in section~\ref{sec:BSM}, we assume the presence of these two operators only.

In figure \ref{fig:(c_W,c_B)-plot} we show the 1-sigma reach from our CLIC analyses in different channels and their combination. 
For comparison we show the direction along which $\hat S$-parameter constraints would lie. Our analysis shows that the CLIC analysis is complementary to a precise $Z$ pole analysis, and that, for the latter to be able to provide improved information, it should reach a precision on $\hat S$ of order $10^{-5}$, in the normalisation of Ref.~\cite{Barbieri:2004qk}, corresponding to $c_B+c_W\sim 2\times 10^{-3}$.

\begin{figure}[h]
\centering
\includegraphics[width=0.6\textwidth]{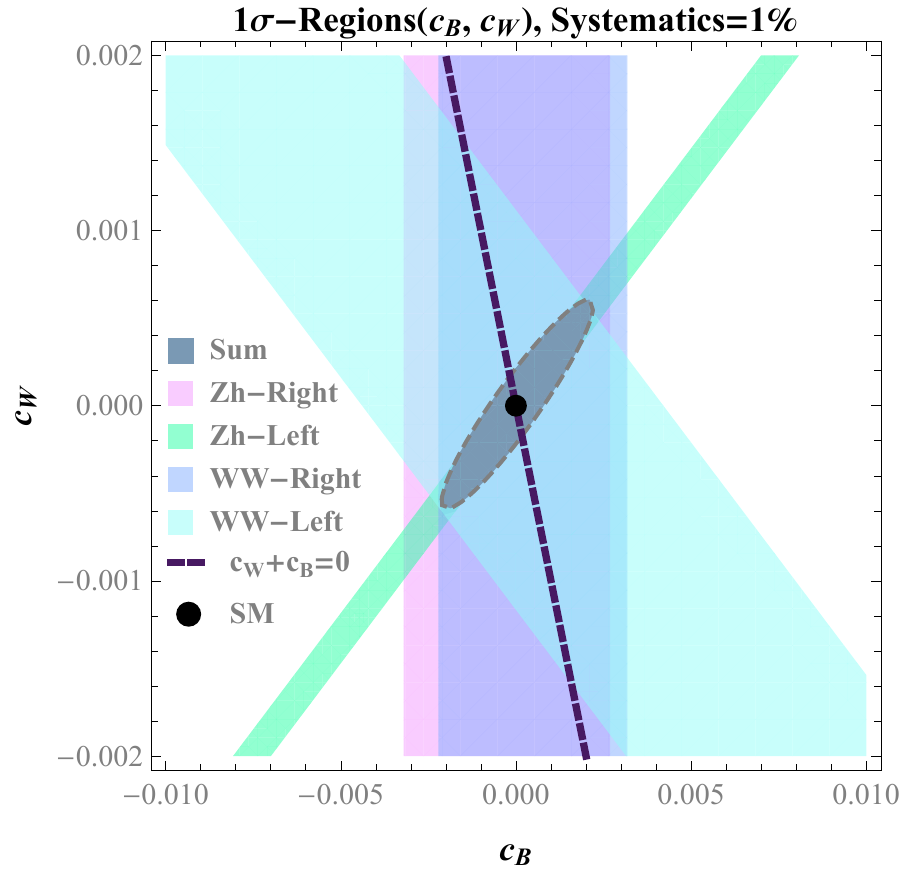}
\caption{\it Probed BSM-directions and their constraints. The shaded areas are derived singling out one process at time, while the solid one is the bound from the summed distribution. The purple line indicates the direction $(c_W+c_B)=0$, studied at LEP.}
\label{fig:(c_W,c_B)-plot}
\end{figure}

\section{Conclusion}
In this work, the high-energy indirect reach on BSM effects in Dibosons production at CLIC and ILC has been assessed, using an EFT approach to interpret the results in a model independent way.

We have divided BSM effects into two classes, those that affect the physics of transverse polarisations and those that affect the longitudinal ones and, by the equivalence theorem, also the Higgs sector.
A study of the transversely polarized $W$'s has been used to explore BSM deviations encoded in the operator $\mathcal{O}_{3W}$, while longitudinal bosons are used to probe new physics in the Higgs-sector in the form of the operators $\mathcal{O}_W$ and $\mathcal{O}_B$. All these operators can be generated in sensible UV-models.

Our results, summurized in tables \ref{tab:Transverseresults}, \ref{tab:Longitudinalsresults}  and \ref{tab:Zhresults} are derived from the analysis of realistic  ILC and CLIC simulations.  We have designed dedicated searches to improve the BSM sensitivity to the above operators.
In particular, the interference in the transverse components has been enhanced using the techniques described in Ref.\cite{Panico:2017frx}, and the possibility of polarizing the beam has been exploited to reduce the background and  improve the sensitivity on longitudinally polarized events.

Despite the fixed center-of-mass energy of lepton colliders, ISR and brehmstrahlung effectively produce a spread in the incoming spectrum that in practice renders the analysis similar to the LHC one. This is manifest in the reconstruction of leptonically decaying $W$-bosons, where the incertitude of the initial momentum along the beam axis translates into an ambiguity in the reconstruction of the azimuthal angle, \eq{eq:amb2}, that makes the analysis blind to CP-odd effects.
Nevertheless, for CP-even effects in the transverse polarizations, our dedicated analysis shows a net improvement, by approximately a factor of 2 with respect to inclusive analyses; this gain is roughly equivalent to a factor of 4 in the effective luminosity.


Our results are interpreted  in terms of a wide class of well-motivated BSM models. For new physics in the transverse polarizations we consider  both weakly-coupled (Figure \ref{fig:transverse}) and strongly-coupled scenarios (Figure \ref{fig:transverse2}), both of which produce a correlated signal (of different relative size) in dibosons and Drell-Yann processes. Quite generally, DY processes offer a better possibility for new discoveries in both cases, yet dibosons still represent an important probe for precision studies. Both processes are sensible to weakly coupled loop-induced effects, if the mass of the new states is not  far above  CLIC direct reach, and if they belong to large representations of $SU(2)_L$.

On the other hand, in the context of longitudinal polarizations, we have compared our study with a synthetic analysis of $Zh$ associated production. The latter has been shown to be generically more competitive when  the effect of a single operator is taken into account. Nevertheless, $WW$ provides complementary as well as additional information when effects from both $\op_W$ and $\op_B$ are included, as shown in Fig.~\ref{fig:(c_W,c_B)-plot}.

\subsection*{Acknowledgements}
We thank Roberto Contino, Sandeepan Gupta, Marc Riembau, Philipp Roloff and Andrea Wulzer for interesting discussions. This work is supported by the Swiss National Science Foundation under grant no. PP00P2-170578.

\bibliography{bib}

\end{document}